\documentclass[%12pt,  
twocolumn,
]{revtex4-2}
\usepackage[draft]{graphicx}
\usepackage{xcolor}
\usepackage[colorlinks]{hyperref}
\usepackage{siunitx} % allows $_\text{}$
\usepackage[normalem]{ulem}

\newcommand{\mm}{~mm$^{-1}$ }
\newcommand{\intens}{~mW/cm$^2$}

\begin{document}
	\title{{Effective interaction dynamics of Rydberg excitons}}  
	\date{\today}
	
	\author{B.~Panda}
	\affiliation{Experimentelle Physik 2, Technische Universit\"at Dortmund, 44221 Dortmund, Germany}
	\author{J.~Heck\"otter}
	\affiliation{Experimentelle Physik 2, Technische Universit\"at Dortmund, 44221 Dortmund, Germany}
	\author{S.~Siegeroth}
	\affiliation{Experimentelle Physik 2, Technische Universit\"at Dortmund, 44221 Dortmund, Germany}
	\author{M.~Harati}
	\affiliation{Experimentelle Physik 2, Technische Universit\"at Dortmund, 44221 Dortmund, Germany}
			\author{M.~Aßmann}
	\affiliation{Experimentelle Physik 2, Technische Universit\"at Dortmund, 44221 Dortmund, Germany}

\begin{abstract}
Rydberg states of excitons in Cu$_2$O are known to interact strongly with  other carriers in their  surrounding such as other excitons, free charge carriers and static charged impurities. 
In cw-experiments, discriminating the individual processes is notoriously challenging. 
Here, we use a time-resolved two-color pump-probe setup  and investigate  the temporal evolution of Rydberg excitons with a  resolution of 15~ns. 
The dynamic response of the Rydberg system to pump pulses with different energies is analyzed and the elementary processes that constitute the complex decay dynamics are compared among the different excitation scenarios. 
Our analysis reveals mainly four different timescales on which physical processes occur. 
\end{abstract}

	\maketitle

\section{Introduction}

The semiconductor material Cu$_2$O is well known as a host-material for Rydberg excitons, i.e. bound electron-hole pairs excited to high quantum numbers $n$.  
So far, Rydberg excitons up to $n=30$ have been observed~\cite{kazimierczukGiantRydbergExcitons2014c, heckotterExperimentalLimitationExtending2020, versteeghGiantRydbergExcitons2021}. 
{Recently, Rydberg excitons were shown to give rise to strong Kerr-type nonlinearities for  visible light~\cite{zielinska-raczynskaNonlinearOpticalProperties2019,morinSelfKerrEffectYellow2022} as well as in the microwave range~\cite{gallagher_microwave-optical_2022, pritchettGiantMicrowaveOptical2024}. Embedded in a cavity~\cite{orfanakisRydbergExcitonPolaritons2022, makhoninNonlinearRydbergExcitonpolaritons2024}, these states become promising candidates for a variety of technological applications~\cite{waltherGiantOpticalNonlinearities2018a}. }

{In highly-excited states the average electron-hole separation reaches the $\mu$m-range. The resulting   dipole moments lead to giant exciton interactions which manifest in an  interaction-induced  nonlinear optical response of the Rydberg system already at low light intensities. }

Among these interactions, researchers identified  strong and long-range dipole-dipole interactions between individual Rydberg excitons 
that result in a phenomenon called Rydberg-blockade~\cite{waltherInteractionsRydbergExcitons2018a, heckotterAsymmetricRydbergBlockade2021,delteilPolaritonBlockadeConfined2019}:
The presence of a Rydberg state shifts the resonance energy of a second Rydberg state out of the excitation linewidth.  This prevents the excitation of other excitons within a certain blockade volume and leads to a reduction of  absorption at the exciton resonance. 
Rydberg blockade may create strongly correlated states and is considered as a key concept for technological applications in quantum information processing~\cite{lukinDipoleBlockadeQuantum2001,saffmanQuantumInformationRydberg2010a}.

Furthermore, Rydberg excitons  show {strong interactions  with electron-hole plasmas 
~\cite{semkatInfluenceElectronholePlasma2019, stolzScrutinizingDebyePlasma2022}: }
The plasma lowers the band gap of the material already at densities on the order of 0.01~$\mu$m$^{-3}$ which leads to a Mott transition of highly-excited  states {and renders Rydberg excitons efficient sensors for ultra-low plasma densities.} Accompanied by a linewidth broadening~\cite{stolz_scattering_2021}, a plasma  leads to a reduction of exciton absorption. 

{Moreover,  Rydberg excitons are affected by electric stray fields of charged impurities in the crystalline environment, even in high-quality  crystals~\cite{krugerInteractionChargedImpurities2020}. In close vicinity to a charged impurity, the Rydberg exciton energies are Stark-shifted away from the bare exciton resonance  which hinders the excitation of excitons by a spectrally narrow laser beam in the surrounding of an impurity.} 
Similar to Rydberg- and plasma blockade, the {interaction with} impurities results in an attenuation of absorption.  
Recently, it was shown how to {employ the Rydberg exciton-impurity interaction} to neutralize charged impurities  which in turn significantly reduces their stray electric fields and enhances the exciton absorption a process called purification~\cite{bergenLargeScalePurification2023, heckotter_neutralization_2023}. 
However, optical excitation may also photoionize neutral impurities which counteracts purification. 
	
The strong mutual interactions between Rydberg excitons  and other Rydberg states, an electron-hole plasma or charged impurities are of  both  fundamental and technological interest. 
An inevitable experimental tool  to investigate these interactions is optical  pump-probe spectroscopy, where a pump laser is used to excite a certain state of the system and a second laser is used to probe the response of a Rydberg exciton~\cite{heckotterAsymmetricRydbergBlockade2021}. 
However, in experiments based on optical excitation, a complex mixture of  Rydberg excitons, an  electron-hole plasma and charged impurities is  created simultaneously. 
Exciton-exciton, exciton-plasma and exciton-impurity interactions all induce a change of absorption that adds up to an altogether complex spectral signature at the exciton resonance.  This renders  a discrimination of the different interactions non-trivial. 
Common strategies to single out one particular interaction make use of  characteristic scaling laws inherent to the underlying interaction potentials~\cite{heckotterAsymmetricRydbergBlockade2021} or isolate particular  intensity regimes at which one effect dominates over others~\cite{stolzScrutinizingDebyePlasma2022}. 

Here we  exploit the different time scales that underlie the interactions in order to discriminate  different interaction mechanisms of Rydberg excitons. Using a pulsed excitation scheme and a time-resolved detection with a temporal resolution of 15~ns, we can separate the individual interaction processes. 
We measure the dynamic response of a Rydberg exciton absorption line to a pump beam at different excitation energies. 

The time-resolved transients carry information about the different processes involved in the interaction and allow us to distinguish and separate physical processes that act on different timescales. 
The observed processes are analyzed and compared for the different excitation energies to shed light onto the complex interplay of the optically excited entities. 
Moreover, we identify the corresponding intensity regimes, within which the different processes occur.

\section{Experiment}
\subsection{Setup}
\begin{figure}
	\includegraphics[draft=false,width=\linewidth]{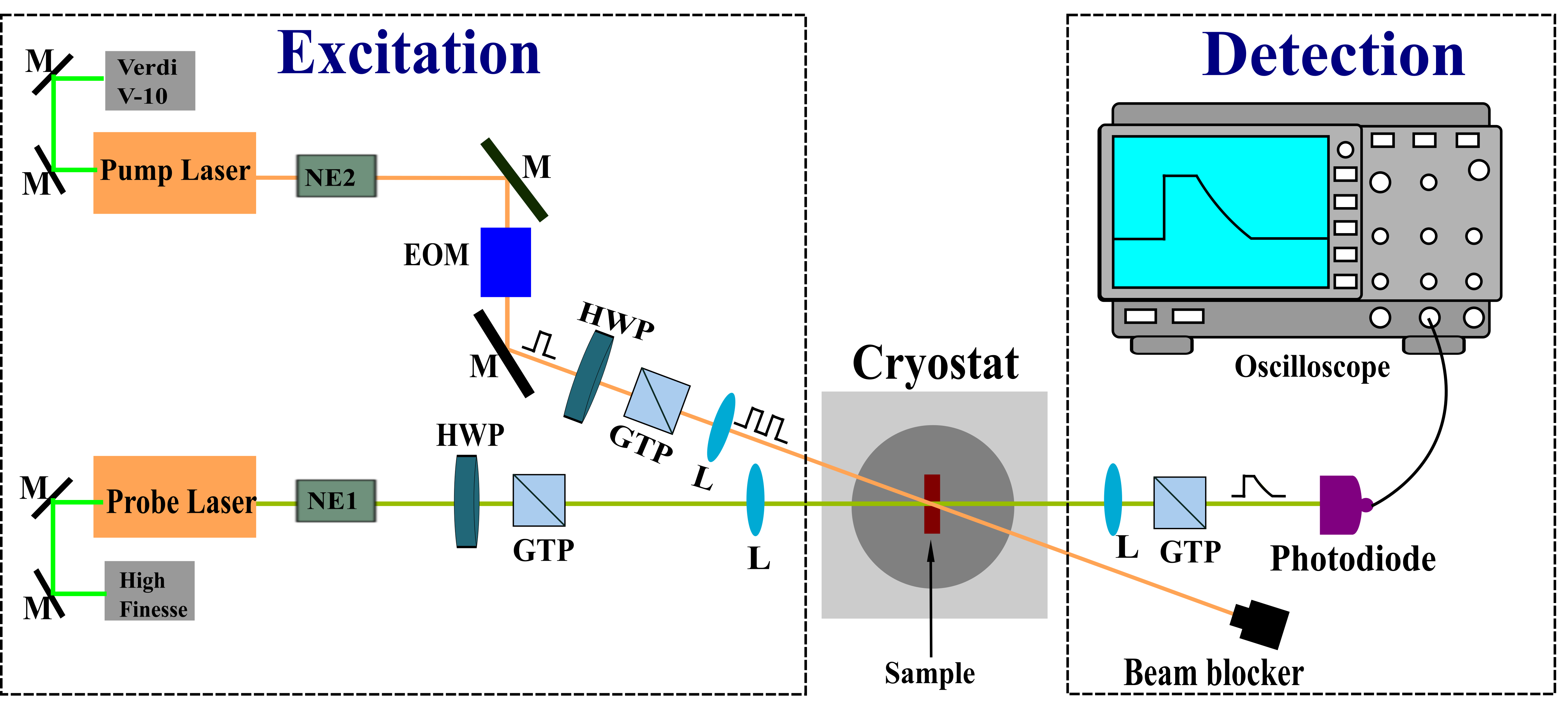}
	\caption{Schematic representation of experimental setup: Modulated pump light periodically induces changes in a cryostat-housed sample which is probed using CW probe laser.
		The dynamic response is detected by a fast photo-diode and an oscilloscope. Abbreviations: NE: noise eater; M: mirror; EOM: electro-optic modulator; HWP: half wave plate; GTP: glan-taylor prism; L: lens}
	\label{fig:Schematic experimental setup}
\end{figure}

A schematic drawing of the experimental setup is shown in Fig.~\ref{fig:Schematic experimental setup}. 
A high-quality natural $Cu_{2}O$ sample of thickness $d=34~\mu$m is used for the experiment~\cite{kazimierczukGiantRydbergExcitons2014c}. It is placed in  a cryostat surrounded by liquid helium providing a temperature of 1.3~K. 
We employ two tunable narrow linewidth dye lasers (Sirah Matisse DS) in a  two-colour pump-probe setting. The lasers each provide a linewidth of around 10~neV which is significantly narrower than any individual Rydberg exciton line.
Within a single run of the experiment, both lasers are locked to particular energies within the absorption spectrum in order {to study the effective interactions that can be traced back to excitation  of well-defined exciton states.} In different runs of the experiment the laser energies are varied to obtain a systematic overview of the interactions between different states. 
The probe laser is employed in continuous wave (cw) mode, while the pump laser beam is temporally modulated. 
The modulation of the pump beam is performed using an electro-optic modulator (EOM) at a frequency of 10~kHz and with $2\%$ duty cycle. The frequency and duty cycle are chosen based on the longest temporal dynamics we are interested in, which fully decay after about 80 $\mu s$.
The profile of the pump laser pulse is shown as a blue line in Fig.~\ref{fig:PumpInduced decay example}(a). The rise and fall times are as fast as 15~ns, which defines our temporal resolution, see inset. 
The spot diameter  of the pump laser at the sample position is 300~$\mu$m ({measured at full-width-half-maximum}). It is three times larger than that of the probe laser which amounts to 100~$\mu$m ({full-width-half-maximum}). The pump spot fully covers the probe spot, which is ensured by maximizing the pump-induced change of probe signal on an oscilloscope. 
Probe powers of 200~nW and 500~nW are used, resulting in cw probe intensities of 1.5\intens{} and 4\intens{}. For conversion between power and intensity, see Appendix~\ref{appendix1}.  
Both laser beams are cross-polarized in order to filter stray light from the strong pump beam. 
A Glan-Taylor prism in the probe beam path placed behind the cryostat filters the pump beam stray light while it also ensures maximum transmission of the probe laser light.\\

On the detection side, we use a fast avalanche photo diode (Thorlabs, APD430A2) with 400\,MHz bandwidth to measure the transmitted intensity of the probe laser. The detected probe light intensity is monitored and stored using an digital oscilloscope with a temporal resolution of 0.5~ns. We obtain three different types of signals: one without any laser light ($S_{bkg}$), which quantifies the background noise level arising from the surrounding room light and diode bias. The second one is proportional to the probe laser transmission in absence of the pump laser, $T_\text{pr}=T_0\exp(-\alpha_\text{pr} d)$, where $T_0$ is the transmitted intensity in front of the cryostat and $\alpha_\text{pr}$ the absorption coefficient seen by the probe laser. This signal acts as reference. The third one measures the change in probe laser transmission induced by the pump pulse, $T_\text{pr,pu}=T_0\exp{(-\alpha_\text{pr,pu}d)}$.

%%%%%%%%%%%%%%%%%%%%%%%%%%%%%%%%%%%%%%%%%%%%%%%%%%%%%%%%%%%%%%%% 
Throughout this paper, we discuss the corresponding change in the absorption coefficient of the probe laser induced by the pump laser and define it as $\Delta \alpha=\alpha_\text{pr,pu}-\alpha_\text{pr}$. 
It results from the relative transmission $T_\text{pr,pu}/T_\text{pr}$, after subtraction of the corresponding background signals: 
\begin{equation}
\Delta\alpha=%-ln\left(\frac{T_\text{pr+pu}-S_\text{bkg}}{T_\text{pr}-S_\text{bkg}}\right)\Big/d=
-\ln\left(\frac{T_0\exp{(-\alpha_\text{pr,pu}d)}}{T_0\exp{ (-\alpha_\text{pr} d) }}\right)\Big/d= \alpha_\text{pr,pu}-\alpha_\text{pr}  \ . 
\label{Formula for Delta Alpha}
\end{equation}

Positive values of $\Delta\alpha$ indicate a pump-induced  increase of probe absorption, a phenomenon that is referred to as purification~\cite{bergenLargeScalePurification2023, heckotter_neutralization_2023}. A common origin for the appearance of purification is the neutralization of charged impurities via Rydberg excitons, which in turn reduces stray electric fields in the crystal. 
Negative values of $\Delta\alpha$ instead indicate a decrease of absorption, known from effects such as Rydberg blockade~\cite{heckotterAsymmetricRydbergBlockade2021}- or plasma-blockade~\cite{stolzScrutinizingDebyePlasma2022}. Note that also optically induced photoionization of impurities which may loosely be described as  purification in reverse can lead to a reduction of probe absorption, see Sec.~\ref{sec:Results} for more details.  All effects may compete as will be pointed out in the next section.

 %%%%%%%%%%%%%%%%%%%%%%%%%%%%%%%%%% EXAMPLE ANALYSIS
 \section{Results}\label{sec:Results} 
 \subsection{Pump pulse response}
 \begin{figure*}
 	\includegraphics[draft=false,width=\linewidth]{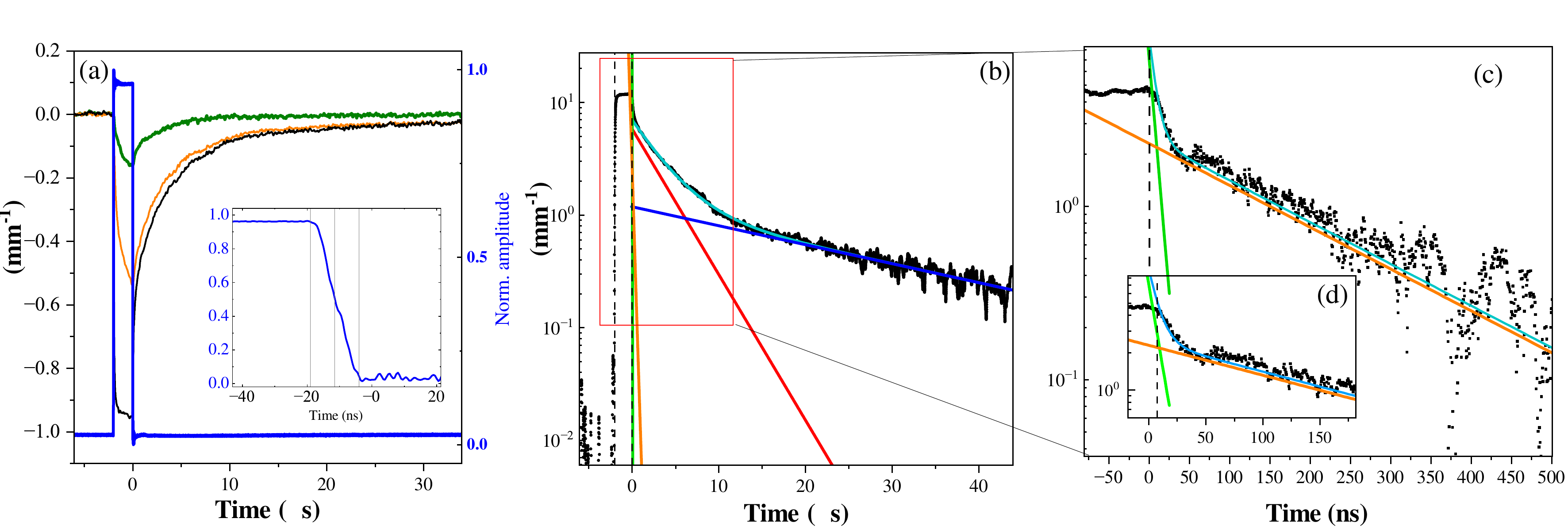}
 	\caption{(a) Probe beam absorption resonant with $n_{probe}=10P$ upon excitation by a pump beam for 2~$\mu$s as a function of time for three different pump intensities of 17 mW/cm$^2$ (green), 170 mW/cm$^2$ (orange) 2673 mW/cm$^2$ (black).   The pump laser photon energy is set to 2.117957~eV, exciting $1S$ excitons indirectly.
 		The blue curve shows the time-profile of the pump pulse created by the EOM. The rise and fall times are about 15~ns (see inset). 
 		(b) Black curve from panel (a) but on logarithmic y-axis to visualize the different decay channels. The transient is composed of four decays, from which two  decay on a $\mu$s timescale (blue and red line). The data is smoothed for better visualization. 
 		Start and end of the pump pulse are indicated by the vertical dashed lines.
 		(c) Zoom into the first 500~ns after the pump pulse ends. {The long decays (blue and red in panel (b)) are subtracted. On this time scale, the two fast decays dominate with decay times in the ns range (orange and green).} Note the {orange  one} is shifted for better visibility. 
 		(d) Zoom into the first 150~ns after the pump pulse ends. Around 50~ns an oscillation is observed, which is neglected (see text).}
 	\label{fig:PumpInduced decay example}
 \end{figure*}

The continuous-wave low-intensity probe laser is used to monitor {the absorption at the $10P$ exciton resonance.} At ultralow intensities the probe beam  creates a small finite density {of $10P$ Rydberg excitons} that depends on the oscillator strength and exciton lifetime of the particular state. Relaxation of Rydberg states and direct absorption into the phonon background may also create a certain density of $1S$ excitons~\cite{schonePhononassistedAbsorptionExcitons2017}. 
{Moreover, the laser excitation does not just create carriers, but also determines the dynamic equilibrium between charged and neutral impurities in the crystal as pointed out in  Ref.~\cite{bergenLargeScalePurification2023}:} Any illumination by a laser with photon energies above the impurity binding energy can drive impurities from a neutral to a charged state by photoionization. Rydberg excitons, in turn, can neutralize charged impurities and drive them to the neutral state with a certain capture rate mainly governed by the quantum number of the exciton. The steady state of the impurity system in case of cw-illumination is then given  by the ratio between photoionization- and neutralization rates. 

When the pump pulse arrives, it drives the system out of the dynamical equilibrium state determined by the probe laser. 
{The pump-induced changes may include both blockade effects introduced by the created carriers and modified ionization and neutralization rates of the impurities.} 
The carriers  excited by the pump laser either increase or decrease the probe laser absorption. 
An example curve is given in Fig.~\ref{fig:PumpInduced decay example}(a) for three different pump intensities.  \\
{In this particular case, the pump laser photon energy  is set to 2.117957~eV, exciting $1S$ excitons indirectly via the phonon absorption band.} 
The pump laser induces a decrease of probe laser absorption, resulting in negative values for $\Delta \alpha$. 
As soon as the pump pulse ends, the probe absorption relaxes back towards the non-perturbed state within several tens of $\mu$s. 
In order to unravel the different physical processes that occur during the presence of the pump pulse, we focus on the different relaxation timescales on which the system reverts back from the new to the old steady state. We discuss  the pump power dependence of both the exponential decay times $\tau$  and the amplitudes $\Delta\alpha$ associated with the different physical processes. 

Figure~\ref{fig:PumpInduced decay example}(b) {shows the  experimental results for the highest pump power, but} on a semilogarithmic scale. In this chart, up to {four} exponential decay times become visible, which are indicated by  solid lines. 
Two processes, depicted by the blue and red lines, show long time scales $\tau_3$ and $\tau_4$ on the order of $\mu$s. The sum of both is shown as the cyan line. 
Figure~\ref{fig:PumpInduced decay example}(c) {shows the transient after subtraction of the two long decays within the first 500~ns.  }
Two fast processes become visible that decay on a nanosecond timescale, depicted by the orange and green lines. 
The process indicated in green shows a decay as fast as the temporal resolution of 15~ns, while the remaining one is on the order of 100~ns.

We note that some of the measurements reveal an additional oscillatory feature within the first 100~ns after the pump pulse. 
An example is shown in Fig.~\ref{fig:PumpInduced decay example}(d).
These oscillations may stem from residual electromagnetic stray fields received by the photodiode that are caused by driving the EOM with fast rise- and fall times. 
So far, their impact on the transients in this short time range cannot be ruled out completely and may cause the observed dip. 
Therefore, we refrain from interpreting this as an additional physical process and  neglect it in the analysis throughout this manuscript.

\subsection{Variation of pump laser energy}\label{subsec:Variation of pumping laser energy} 
 \begin{figure*}
\includegraphics[draft=false,width=\linewidth]{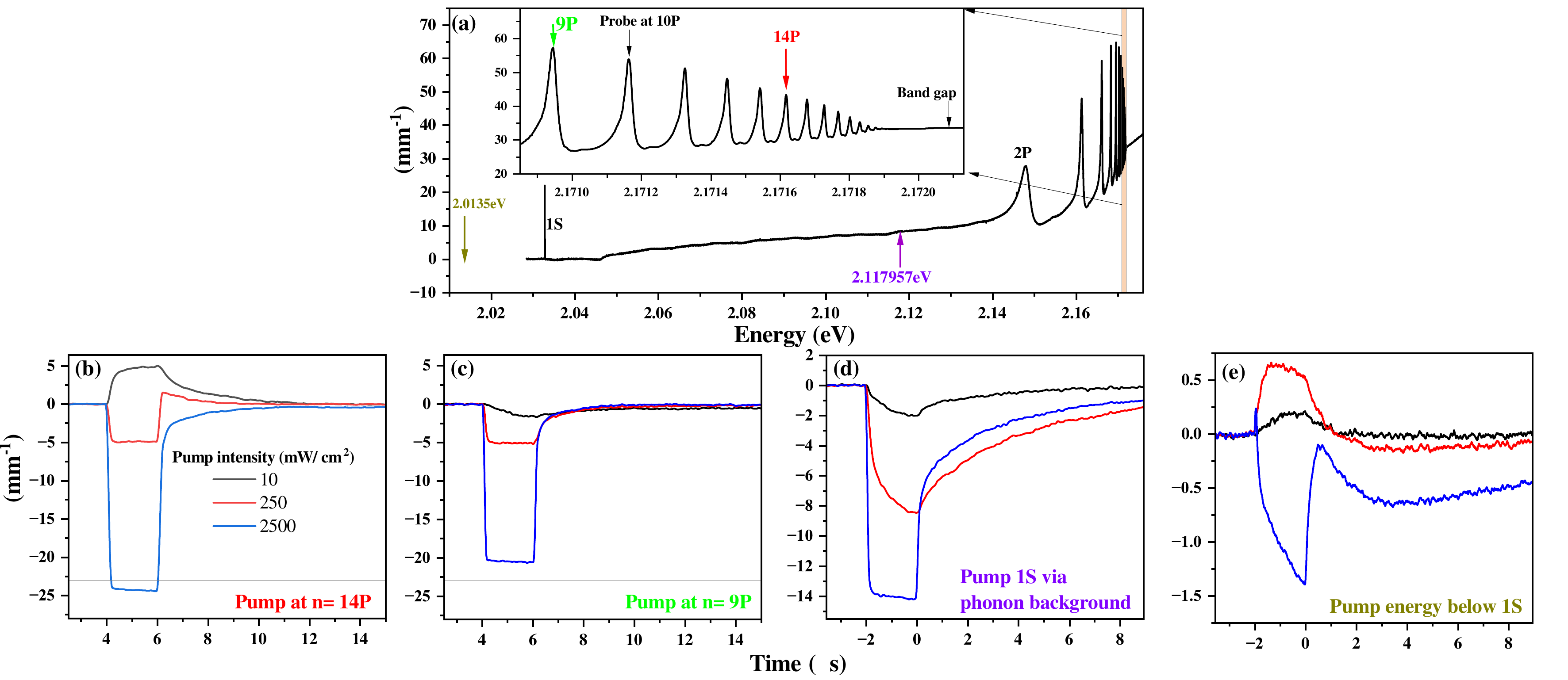}
\caption{(a) Linear absorption spectrum of the yellow exciton series starting from the $1S$ exciton at 2.0328~eV to an energy of 2.176~eV above the band gap. Data taken from Ref.~\cite{heckotterthesis, stolzCoherentTransferMatrix2021}.  At a phonon energy of 13.6~meV above the $1S$ energy, a broad phonon continuum starts. On top of it, distinct exciton resonances become visible from 2.14~eV onwards. The four energies at which the pump laser is fixed during measurement, are at $n=14P$, $n=9P$, 2.117957eV and 2.0135eV, which are marked in red, green, purple and brown colours respectively. The probe laser position at $n=10P$ is indicated in black colour. (b), (c), (d) and (e) demonstrate the pump laser-induced  change of probe laser absorption for each pump energy and three different pump intensities. The data is smoothed for better visibility. }
\label{fig:PumpInducedDecay_4Demo}
\end{figure*}
In the experiment, the probe laser is fixed at the resonance $n_{probe}=10P$ and the pump laser energy $E_{pump}$ is successively tuned to four different energies that are discussed qualitatively in the following.  
As a reference, Fig.~\ref{fig:PumpInducedDecay_4Demo}(a) shows the absorption spectrum of the whole yellow exciton series.  
For clarity, the spectral positions of the pump and probe  lasers are indicated in the spectrum by the colored vertical arrows.  

%%%%%%%%%%%%%%%%%%%%%%%%%%%%%%   DESCRIBE FIGURE WITH DETAILS
First, we consider the case where the pump laser excites the $14P$ Rydberg exciton resonantly (red arrow in (a)). 
Hence, we have the case $n_\text{pump}>n_\text{probe}$. 
Second, the pump laser excites the $9P$ Rydberg exciton resonantly (green arrow) and we consider $n_\text{pump}<n_\text{probe}$ in this and the following case.
Third, the pump laser excites the $1S$ exciton indirectly via the broad phonon background at 2.117957~eV (violet arrow). This excitation scheme is comparable to the one studied for cw-excitation in Ref.~\cite{stolzScrutinizingDebyePlasma2022}. 
Here, after Auger-decay of $1S$ excitons, an electron-hole plasma is formed~\cite{kavoulakis_auger_1996,ohara_auger_1999,jang_auger_2006} which is the main source of changes to the optical properties of Rydberg excitons in this excitation scenario. For the final setting, the pump laser photon energy is set to 2.0135~eV, an energy below the $1S$ exciton line (brown arrow). At this energy no absorption is expected a priori. 

The time-resolved absorption change $\Delta \alpha$ of the probe beam resonant with the $10P$ exciton for all four cases is shown in Fig.~\ref{fig:PumpInducedDecay_4Demo}(b)-(e), for three different pump intensities.  
We start with a discussion  of the pump beam being resonant with $14P$. 
The pump laser induces an enhancement of absorption at the lowest pump intensity of 10\intens{}, as shown by the black curve in Figure~\ref{fig:PumpInducedDecay_4Demo}(b). {Such an increase in absorption stems from the neutralization of intrinsic charged impurities by Rydberg excitons as described in Refs.~\cite{bergenLargeScalePurification2023, heckotter_neutralization_2023}.}
At the end of the pump pulse, the absorption reverts back to its initial value on a $\mu$s time scale. Such long time scales that exceed the lifetimes of Rydberg excitons clearly indicate the participation of impurities in the associated process. 
% HIGHER intensityS%%%%%%%%%%%%%%%%%%%%%%%%%
At 250\intens{} the probe absorption instead decreases due to the presence of the pump pulse. 
When the pump pulse is gone, the absorption change decays on a {nanosecond time scale, which is much faster than the one observed at lower intensity.} Consequently, species with lifetimes on the nanosecond scale, such as $14P$ Rydberg excitons, may contribute to this process. 
Interestingly, the change in absorption even changes %sign,  growing to positive values from where it decays back to the initial steady state on a $\mu$s-long time scale. 
{sign after the end of the pump pulse. Initially the change in absorption quickly grows towards positive values within a few tens of nanoseconds and then decays back to the initial steady state on a $\mu$s-long time scale.}
At the highest pump intensity of 2500\intens{}, the pump induces a reduction of the probe laser absorption 
as well. {The absorption drops by almost 25~mm$^{-1}$, which is close to  the initial bare absorption of the $10P$ line above the phonon background (horizontal line). Note that an unavoidable pump laser offset is subtracted here for better visibility. } Hence, the probed exciton absorption line of the $10P$ is fully bleached. After the pump pulse has ended, the system reverts back to its initial state, whereby the major part of $\Delta \alpha$ of around 20~mm$^{-1}$ decays rather fast, followed by  slower processes on the $\mu$s scale. 
Remarkably, at this high pump power  all processes have negative signs, even the slow decays which show positive signs at low pump powers. 

%%%%%%%%%%%%%%%%%%%%%%%%%%
Figure~\ref{fig:PumpInducedDecay_4Demo}(c) shows the transients for pumping $9P$ excitons. %Hence, we consider $E_\text{pump}<E_\text{probe}$. 
For all pump intensities, the pump induces solely a reduction of probe laser absorption. 
After the end of the pump pulse with highest intensity (blue), the major part of the absorption reverts back again on a fast time scale directly. 
For all three pump intensities, slower processes with smaller amplitudes can be observed for microseconds after the end of the pump pulse. 
Since the $9P$ excitons are resonantly excited,  here the fast decay  may stem from Rydberg blockade since a huge density of $9P$ Rydberg excitons is created. 
{For $n_\text{pump}<n_\text{probe}$ also the impurity system may contribute to the reduction of absorption 
since the presence of the pump beam increases the impurity ionization rates relative to the capture rates compared to when only the probe beam is present~\cite{bergenLargeScalePurification2023}.} 

{Moreover, also an electron-hole plasma  created via Auger-decay of $1S$ excitons can, in principle, contribute to the reduced absorption in Figs.~\ref{fig:PumpInducedDecay_4Demo}(b) and (c). For these excitation schemes, $1S$ excitons may form via phonon-assisted absorption or after relaxation from $16P$ and $9P$ Rydberg states.}
We assume that purification and the presence of a plasma may contribute to the slower processes we observed. 

%%%%%%%%%%%%%%%%%%%%%%%%%%%
Exciting $1S$ excitons via the phonon background also results  solely in negative amplitudes of $\Delta \alpha$ (Fig.~\ref{fig:PumpInducedDecay_4Demo}(d)). For the lowest pump intensity (black curve), the dynamical change in absorption is comparable to the case in panel (c). 
At medium intensities (red), however, the indirect excitation of $1S$ excitons causes a change in alpha of more than 8~mm$^{-1}$.
When exciting $9P$ excitons with the same pump intensity, only a smaller amplitude $|\Delta \alpha| \approx 5$~mm$^{-1}$ is measured (panel (c)), {although the absorption is $\sim$ 6 times higher here.} Note the different scaling of the ordinates. The amplitudes in both cases are discussed further below in Sec.~\ref{plasma}. Also here, a fast decay of the probe laser absorption is observed at the highest pump intensity, although its magnitude is smaller compared to the case in panel (c).

%%%%%%%%%%%%%%%%%%%%
Last, we consider a pump energy below the resonance energy of the $1S$ exciton. This is shown in Fig.~\ref{fig:PumpInducedDecay_4Demo}(e). 
At this energy the material does not show any significant amount of linear absorption (see e.g. Ref~\cite{schonePhononassistedAbsorptionExcitons2017}). 
Hence, we expect no direct excitation of excitons, but photo-ionization of impurities is still possible. 
Remarkably, here we observe a well-defined response of the $10P$ excitons  as well, although the change in $\alpha$ is ten times smaller compared to the other cases. Note the different scale on the ordinate. 
Up to 250\intens{}, we observe an increase of absorption  of about 0.5 mm$^{-1}$. While the process of purification is closely connected to the excitation of Rydberg excitons, this observation is remarkable, since the pump energy is not resonant with any exciton energy.  
Another possible excitation process is the formation of a dilute electron-hole plasma via 2-photon absorption that causes purification directly or after relaxation into high lying Rydberg excitons. 
Moreover, instead of relaxation back to the initial state, we observe an overshoot to negative amplitudes from which the absorption slowly decays back to zero (red curve in panel (e)). 
At the highest pump intensity (blue line), the sign of $\Delta \alpha$ is solely {negative during the presence of the pump pulse and the absorption is far away from saturation.} The decay at the end of the pulse seems to consist of a fast decay, a slow growth and another even slower decay.

%%%%%%%%%%%%%%%%%%%%%%%%%%%%%%%%%%%%%%%%%%%%%%%%%%%%%%%%%%%%%%%%%%%%%%%%%

In order to compare the different excitation cases shown in  Fig.~\ref{fig:PumpInducedDecay_4Demo} more quantitatively, the  transients  are decomposed into their constituents. 
In all excitation scenarios primarily {four} decay channels can be identified in a semilogarithmic representation (Fig.~\ref{fig:PumpInduced decay example}). 
The amplitudes and time scales are  estimated by fitting a double exponential function to the data
 \begin{eqnarray}
\Delta\alpha&= \Delta\alpha_{0} +
A_{1}e^{-{t}/{t_{1}}}+\\
&A_{2}e^{-\left({t}/{t_{2}}\right)}+ A_{3}e^{-\left({t}/{t_{3}}\right)}+ A_{4}e^{-\left({t}/{t_{4}}\right)} \ ,\nonumber
\label{fun:double exponential function}
\end{eqnarray}
where $t$ is the time. 
The amplitude of each process is evaluated at $t_0=0$, which corresponds to the moment the pump beam is switched off. 
The first (second) terms describe the fast decays with amplitude $A_{1}$ ($A_2$) and decay time  $\tau_{1}$ ($\tau_2$) on the order of tens (hundredths) of nanoseconds. 
$A_{3}$ and $A_{4}$ are the decay amplitudes corresponding to the two slower decay processes with time scales defined as $\tau_{3}$ and $\tau_{4}$ on the $\mu$s scale. $\Delta \alpha_{0}$ is an offset accounting for non-zero background noise.% and  $t_{0}=0$ is the time stamp at which the pump pulse ends.
%%%%%%%%%%%%%%%%%%%%%%%%%%%%%%%  FIGURE 4
\section{Amplitudes and decay times}
\subsection{Comparison of pump energies}
 \begin{figure*}
	\includegraphics[draft=false,width=\linewidth]{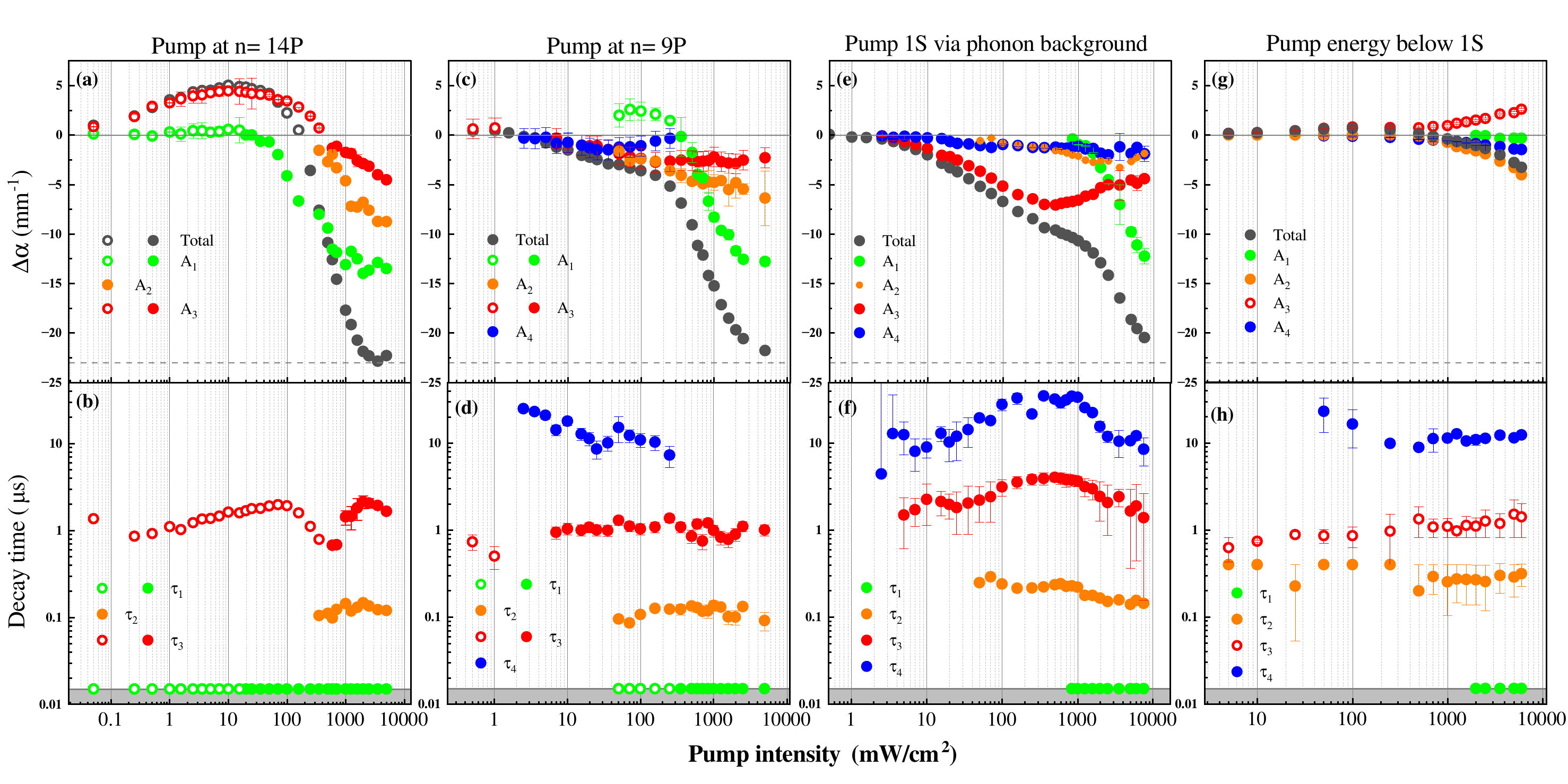}%PumpComparision.png}
	\caption{Upper panels (a), (c), (e) and (g) show the change in absorption coefficient $\Delta\alpha$ as a function of pump laser intensity at four different pump energies. The probe intensity is set to 1.5\intens{} (except in (g), which is measured with 3.8\intens{} ) and probe energy($E_{probe}$) fixed to the resonance $n_{probe}= 10$. {The horizontal dashed lines indicate the maximum possible of $\Delta\alpha=23\pm 1 $mm$^{-1}$) for $n_{probe}= 10$.} 
	Black dots represent the overall absorption changes while green, orange, red and blue colors stand for individual processes whose corresponding time scales are plotted in lower panel (b), (d), (f) and (h). 
}
	
	\label{fig:Pump energy Comparison}
\end{figure*}
The amplitudes and decay times obtained by the fits are displayed in Fig.~\ref{fig:Pump energy Comparison} as a function of pump intensity for all four  pump laser energies. In order to identify the underlying physical processes, we first plot all values corresponding to comparable decay times in the same color. Full symbols indicate negative amplitudes and open symbols indicate positive amplitudes. The panels in the upper row show the amplitudes of the four individual decay channels along with the total amplitude of each transient, evaluated at the end of the pump pulse. 
The horizontal dashed line depicts full bleaching of the $10P$ exciton line, which we determined from cw measurements to amount to $-23\pm 1$ $mm^{-1}$. 
The lower row in Fig.~\ref{fig:Pump energy Comparison} shows the  decay times of the individual  processes. 
The time resolution is indicated by the gray area. 

We first discuss  pumping the $14P$ exciton (Fig.~\ref{fig:Pump energy Comparison}(a) and (b)). 
%%%%%%%%%%%%%%%% color code
%%%%%%%%%%%%%%%%  A1
The decay amplitude and time scale of the fastest decay, $A_1$ and $\tau_1$, are shown in green. The  decay time  is $\leq$ 15~ns and limited by the temporal resolution of the setup. 
This process can be found for all pump intensities, but becomes dominant from 100\intens{} onwards. 
At low intensities it shows up in the transients with positive amplitudes (open dots) and changes sign around 10\intens{} to become negative (full dots).  
{The fast timescale implies that a short-living species contributes to this process which likely are Rydberg excitons to which the pump laser is resonant here. }
%%%%%%%%%%%%%%%% A2 0.1 µs
Around 300\intens{} the second fastest process (orange) with $\tau_2\approx 100$~ns appears in the transients. Its amplitude $A_2$ is negative and its magnitude increases strongly with pump intensity up to 8\mm. 
{Lifetimes of several hundred nanoseconds imply plasma-related particles such as para excitons or free electrons and holes  to play a role. These carriers may stem from an Auger-decay of $1S$ excitons that are indirectly excited via the phonon background.  
%%%%%%%%%%%%%%%%  A3 3 color, sign change
The third process (red) has a decay time  in the range of %$\tau_3=1-3~\mu$s. 
$\tau_3=0.8-2~\mu$s, 
whereas its sign changes from positive to negative with increasing intensity. 
This process dominantly contributes to the observed transient in the low intensity regime up to a pump intensity of 100\intens{}, {where it has positive amplitudes (open symbols) and leads to an increasing probe laser absorption (purification)}. 

At intensities larger than 10\intens{}, $A_3$ approaches zero, indicating purification to become less efficient. 
From 400\intens{} onwards, a process with $\tau_3\approx 1~\mu$s is observed, but with negative amplitude meaning reduction of absorption (full red symbols). 
The similarities of the timescales imply that this process is related to the dynamics of the impurities, but there are no unambiguous experimental signatures supporting this assignment at this point. 
We note that indications of a process on time scales of 10~$\mu$s can be observed in case of excitation of $14P$ excitons as well, but the amplitudes are too small to evaluate them. 
All  processes add up to the total amplitude of about $\Delta \alpha=-23$\mm (black) at the highest pump intensity of 5000\intens{}. %$A_2$ even overcomes the fast process of $A_1$ at high intensities of . 

%%%%%%%%%%%%%%%%%%%%%%% Pump 9P NEU
When pumping the $9P$ state (Fig.~\ref{fig:Pump energy Comparison}c) the fastest process corresponding to $A_1$  is observed to be positive (open green dots) at intermediate intensities, from 50\intens{} to 250\intens{}. 
While the total amplitude (black) is still negative here, the positive amplitude of $A_1$ counteracts the three other processes. 
So far, processes with positive amplitudes were always observed to have slower decay dynamics, e.g. in Ref.~\cite{bergenLargeScalePurification2023}.  
{The fast time scale implies a direct screening of charged impurities by   Rydberg excitons, which disappears as soon as the pump laser is gone. }
In the high-intensity regime, from 500\intens{} upwards, $A_1$ becomes negative (full green dots) and increases strongly in magnitude with pump intensity dominantly contributing to the total amplitude. 
{Similar to the case of pumping $14P$ states, this process is related to the presence of  $9P$ Rydberg excitons. These may induce an attenuation of probe laser absorption directly by Rydberg blockade or after Auger-decay into a plasma.   }
%%%%%%%%%%%%%% A2
We observe a decay with $\tau_2\approx 100$~ns here as well.  Its amplitude $A_2$ is negative and grows monotonously with pump intensity up to -6~mm$^{-1}$ at highest pump-intensities. 
{This is in line with an increasing impact of an electron-hole plasma created by long-living ground state excitons.  }
%%%%%%%%%%%%%% A3 A4
Longer processes with timescales around $\tau_3\approx $1~$\mu$s and $\tau_4\approx $10~$\mu$s are observed as well. 
Although their magnitudes do not become larger than  $2.5$~mm$^{-1}$, they contribute significantly to the decay dynamics as they survive up to 50~$\mu$s, see semilogarithmic representation in Fig.~\ref{fig:PumpInduced decay example}(c). 
{The observation of two different slow decaying processes implies the presence of at least two different types of impurity-related processes.} 
Due to the complex interplay of different mechanisms, the total amplitude shows a step-like behavior consisting of a moderate decrease up to  around 100\intens{} and a stronger drop at higher intensities.  It starts to saturate around 5000\intens{} where the probed $10P$ line is fully bleached.

%%%%%%%%%%%%%%%%%% (e)  PUMP 1S NEU
%%%%%%%%%%%%%%%%%%%%A1 A2
In case of pumping $1S$ indirectly (Fig.~\ref{fig:Pump energy Comparison}(e)),  the fast process $A_1$ (bright green) is visible solely in the high-intensity regime from around 1000\intens{} onward. 
$A_1$ dominates the high-intensity regime and is always negative indicating that no purification occurs. 
At this pump energy, mainly $1S$ ortho  and  para excitons are created. 
	{At high pump intensities,  Auger-decay becomes the dominant decay channel of these excitons which also results in the formation of a dilute electron-hole plasma as a byproduct.} 
The time scales {of such Auger-driven processes} correlate with the corresponding exciton lifetimes. 
{The fast timescales observed here imply a strong  Auger-decay rate of  $1S$ ortho excitons with lifetimes of a couple of nanoseconds}. Also, direct plasma creation via two-photon transitions cannot be excluded. 
Para exciton lifetimes, however, are on the order of several hundred nanoseconds and therefore rather comparable to the timescale $\tau_2$. 
%%%%%%%%%%%%%%%%%%%%A3
In the range of low pump intensities, i.e. up to around 500\intens{}, the transient is dominated by the slow process with negative amplitude $A_3${, that we related to impurity dynamics above.} Its magnitude increases up to 8~mm$^{-1}$ at 350\intens{}, whereupon it starts to decrease again. 
{The decrease starts along with the rise of process $A_1$, which implies that free carriers have a strong effect on the impurity system.  }
%%%%%%%%%%%%%%%%%%%%  A4
The long decay $A_4$ is observed up to highest pump intensities, but its amplitude stays rather small and amounts to less than $-3$~mm$^{-1}$.  
The saddling of the total amplitude at intermediate intensities is less pronounced in this scenario compared to pumping $9P$ excitons and saturation of the $10P$ line is not reached even when going to the highest pump intensity of 7500\intens{}. 

%%%%%%%%%%%%%%%%%%%%% pump below 1S 
% pump below 1S 
In Fig.~\ref{fig:Pump energy Comparison}(g) and (h), we show results for a pump energy below the resonance energy of the $1S$ exciton. 
For this pump energy, the total $\Delta \alpha$ changes sign with increasing intensity from positive to negative and reaches  magnitudes up to only $5$\mm at highest pump intensity.  
At low intensities, the decay dynamics are dominated by $A_3$. % on the timescale of $\tau_3=1$~$\mu$s is observed as well. 
$A_3$ has solely positive amplitude indicating purification. This is remarkable, since no direct excitation of free excitons is expected. 
The positive amplitude $A_3$ is counteracted by the other processes in the high-intensity regime from 1000\intens{} onward,  leading in total to an attenuation of the absorption. 
Here, the second fastest process $A_2$ contributes strongly with magnitudes that are comparable to the magnitudes found for the other pump energies.  The process has decay times around 300~ns which is slightly slower compared to the decay times found for the other pump energies. 
{The fact that it is observed at this pump energy and the corresponding timescale imply a formation of a plasma via  2-photon excitation. }
A process  with long decay times  $\tau_4\approx 10-20~\mu$s and a fast decay channel with $\tau_1\leq 15$~ns   become visible but their amplitudes are negligible. 

\subsection{Comparison of total amplitudes for the excitation of $9P$ and $1S$ excitons}\label{plasma}
Next, we have a closer look at the cases of pumping $9P$ excitons and $1S$ excitons via the phonon background, as the direct comparison of their total amplitudes reveals  a non-intuitive behavior at low intensities. 
Therefore, we compare the total amplitudes for these scenarios directly in Fig.~\ref{fig:setup3}(a). 
{Both traces show a nonlinear, step-like power dependence, consisting of a rather weak change in $\Delta \alpha$ at low intensities that flattens at pump intensities around 	100 to 1000\intens{}, depending on the pump energy. At higher intensities $\Delta \alpha$ starts to drop further with a steeper decrease indicating the onset of another process at high intensities that we discuss further below. First we focus on the low intensity regime. }
The absorption coefficient for the pump laser resonant with $9P$ is around 57~\mm, whereas it amounts only to 11~\mm in case of pumping $1S$ excitons via the phonon background at 2.12~eV(see Fig.~\ref{fig:PumpInducedDecay_4Demo}(a)). 
On the contrary, a  larger change in $\Delta \alpha$ {for the probe beam} is observed in the latter case up to 1000\intens{}. 
In order to understand this observation we take a closer look at the individual processes in Fig.~\ref{fig:Pump energy Comparison}(c) and (e). 
For a pump laser resonant with $9P$,  $A_1$ contributes positively to the total amplitude leading in turn to less negative values of the total amplitude compared to  indirect excitation of $1S$ excitons.

\begin{figure}
		\includegraphics[draft=false,width=0.8\columnwidth]{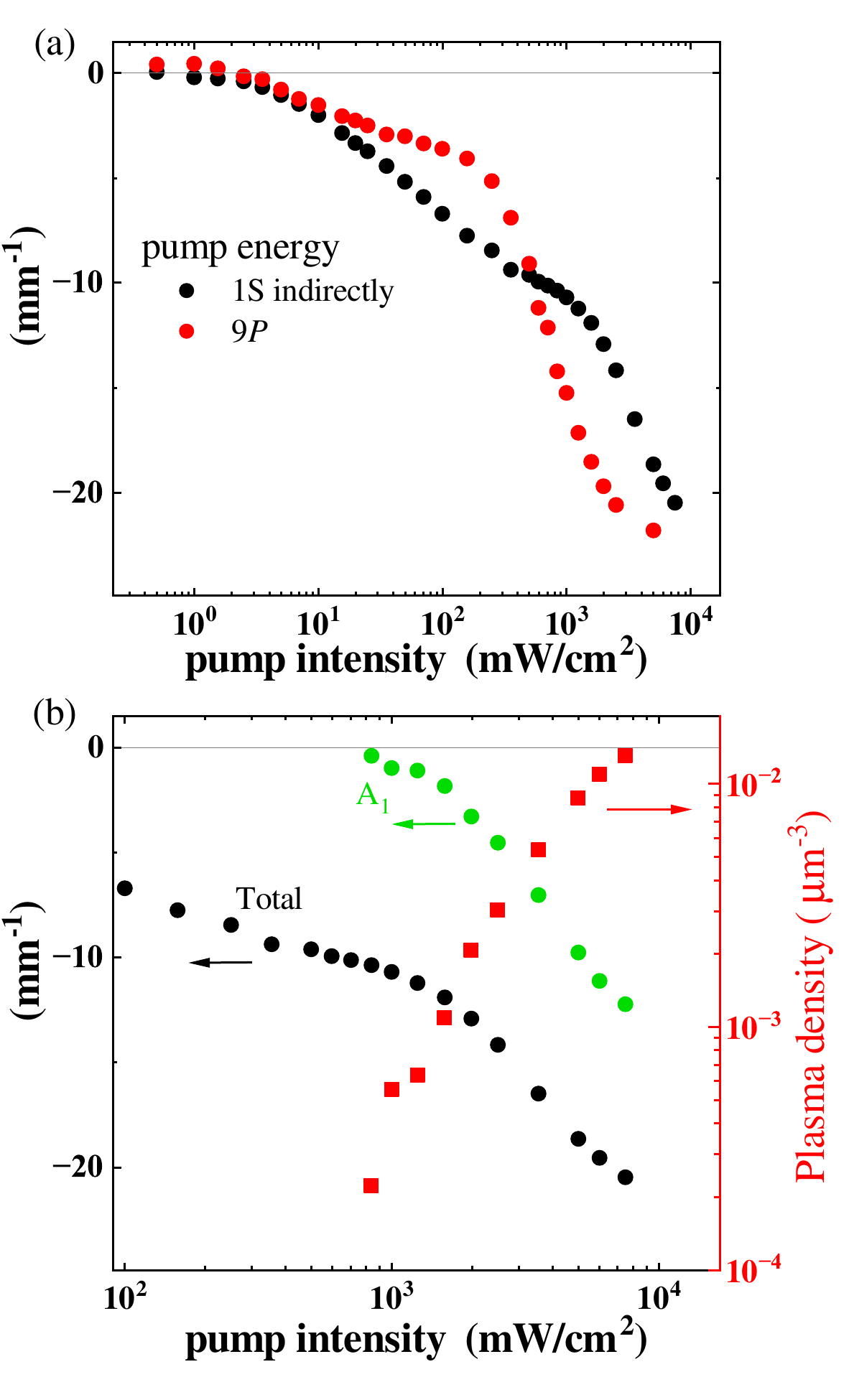}
	\caption{(a) Comparison of total change in $\Delta\alpha$ at the $10P$ exciton as a function of pump laser power when pumping  $9P$ excitons resonantly (red) or $1S$ excitons via the phonon background (black). 
		(b) {Estimation of  plasma densities in  the high intensity regime for pumping $1S$ excitons indirectly.  The amplitude $A_1$ (green, left ordinate) results in  plasma densities (red, right ordinate) between $10^{-4}$ and $10^{-2}$} $\mu$m$^{-3}$.}
  	\label{fig:setup3}
\end{figure}
 
\subsection{{High intensity regime for pumping $1S$ excitons}}\label{A2}
{In order to interpret the results observed for the high-intensity regime in case of pumping $1S$ excitons, we use the observed pump-induced  $\Delta \alpha$ to estimate the effective plasma densities for each pump intensity. The estimation is based on a model presented in  Refs.~\cite{stolz_scattering_2021,stolzScrutinizingDebyePlasma2022}, where  a change in $\Delta \alpha$ is caused by a linewidth broadening of exciton lines by plasmon scattering. This model was developed based on a comparison to cw-data and was found to be valid in the regime of high pump intensities. 
The total linewidth of an exciton surrounded by a plasma is given by
\begin{equation}
	\Gamma_\text{total}=\Gamma_\text{0}(n)+\Gamma_\text{plasma}(n) \ .
\end{equation}
The first term includes phonon scattering and radiative decay. The second term describes the linewidth broadening caused by the plasma
\begin{equation}
	\Gamma_\text{plasma}(n)=C_\text{add}(n-1)^{3.3}
\end{equation}
with $C_\text{add}=3.2\rho_\text{eh}/\sqrt{T_\text{eh}/K}\mu eV \mu m^3$. 
}

The change in absorption of an exciton line is then given by
\begin{eqnarray}\label{linewidth}
	\Delta \alpha&=&\alpha_\text{pr,pu}-\alpha_\text{pr}\\
	%&=&\frac{2}{\pi}O(n)\left(\frac{1}{\frac{1}{\Gamma_\text{total}}-\Gamma_0}\right)\\
	&=&\frac{2}{\pi}O(n)\left(\frac{1}{\Gamma_\text{0}+\Gamma_\text{plasma}(n)}-\frac{1}{\Gamma_0}\right) \ ,
\end{eqnarray}
where $O(n)$ is the oscillator strength of the exciton line with principal quantum number $n$. By rearranging Eq.~\ref{linewidth} we calculate the plasma density $\rho_\text{eh}$ for the amplitude $A_1$, which is the dominant process at high intensities. The densities are shown in Fig.~\ref{fig:setup3}(b) on the right ordinate. They range from $10^{-4}$ $\mu$m$^{-3}$ to $10^{-2}$ $\mu$m$^{-3}$ which agrees with the densities observed in cw-experiments at high pump intensities~\cite{stolzScrutinizingDebyePlasma2022}.

\section{Discussion and Outlook}
In summary, we have experimentally investigated the complex temporal response of Rydberg excitons under various excitation conditions.  We identified four different timescales on which processes appear.  Interestingly, most timescales are observed in all of the excitation scenarios.

%%%%%%%%%%%%%%%%%%%%%%%%%%%%%%%%%% A1
The fastest decay with amplitude $A_1$ is found for all pump energies and intensities. Its observed decay time is limited by the time resolution of our detector. {Therefore, it should not be understood as a single process, but rather as the sum of all relevant processes that we cannot resolve due to limited temporal resolution.} Positive amplitudes are solely found for pump energies resonant with Rydberg excitons (open green dots in Fig.~\ref{fig:Pump energy Comparison}(a) and (c)), pointing towards Rydberg excitons as the relevant species for an increase of absorption. 
At high pump intensities this process invariably shows negative amplitudes, corresponding to a decrease of absorption, and even becomes the main contribution to the total amplitude. 
%%%%%%%%%%%%%%%%%%%%%% pump phonon background
{In Fig.~\ref{fig:setup3}(b), we showed that the amplitude of this process for pumping $1S$ excitons can be explained by plasma densities which are comparable to  densities obtained in  cw-experiments with equivalent pumping conditions which were shown in Ref.~\cite{stolzScrutinizingDebyePlasma2022}. There, paraexciton Auger decay is the dominant source for formation of the plasma. {However, the paraexciton lifetime is on the order of several hundred nanoseconds and we expect any effects originating from the corresponding Auger-generated electrons and holes to decay on a comparable timescale.} Here, the decay associated with $A_1$ is faster than 15~ns which implies that the plasma forms via a different pathway, e.g., Auger decay of orthoexcitons with shorter lifetime.}

%%%%%%%%%%%%%%%%%%%%%%%%%%%%%%%%%% A2
%The second fastest process found in the data
{In contrast, the other fast process $A_2$ shows decay times around $\tau_2=100-300$~ns and also occurs for all excitation energies.} This timescale is way longer and may be related to  a plasma formed by Auger decay of long-living para excitons.

%%%%%%%%%%%%%%%%%%%%%%%%%%%%%%%%%% A3
For all measurements, we observe a slower process arising on the time scale of  $\tau_3=1~\mu$s. Its amplitude $A_3$ may become both positive (increase of absorption) and negative (decrease of absorption), depending on pump power and pump laser energy. 
For positive amplitudes this process is consistent with   purification, i.e. the neutralization of charged defects by capture of an exciton. 
{The dynamical equilibrium model developed for impurities in the presence of Rydberg excitons~\cite{bergenLargeScalePurification2023} predicts an increase of probe absorption due to enhanced purification for $n_\text{pump}>n_\text{probe}$ and a reduction of probe absorption due to enhanced impurity ionization for $n_\text{pump}<n_\text{probe}$. This matches exactly with our observations for $A_3$ for pumping $14P$, $9P$ and $1S$ exciton states, respectively. }
In the latter case, this process even appears to be quite dominant up to intermediate pump intensities. This regime of pump intensities agrees with the "low power range" in Ref.~\cite{stolzScrutinizingDebyePlasma2022}. There, {the observations in that power regime were}  assigned to impurity-related effects and not investigated further. 
{$A_3$ changes sign (Fig.~\ref{fig:Pump energy Comparison}(a)) or gets attenuated (Fig.~\ref{fig:Pump energy Comparison}(e)) as soon as $A_1$ becomes dominant which implies that plasma formation and free carriers also have a strong effect on the impurities and may ionize them effectively.}

%%%%%%%%%%%%%%%%%%%%%%%%%%%%%%%%%% A4
The slowest process with amplitude $A_4$ shows decay times on the order of tens of microseconds. Its amplitude is rather small and always negative, contributing at most  10~\% to the total amplitude. 
This process is observed for pump energies below the $1S$ state as well (Fig.~\ref{fig:PumpInducedDecay_4Demo}(g)/((h)). 
{It likely stems from a long-lived state, such as a particular type of charged impurity. }

\section{Appendix}
\subsection{Calculation of peak intensity}\label{appendix1}
In order to calculate the pump laser peak intensity, we measure the average power $P_\text{avg}$ in front of the cryostat with an accuracy of 0.01~$\mu$W. Losses by reflection of 4~\% at three cryostat windows (six surfaces) and 25~\% at the first sample surface reduce the incoming power by a factor of $\sim0.59$. The peak intensity per pulse is then given via $I_\text{peak} =0.59\frac{P_\text{avg}}{\pi f\tau r_\text{beam}^2}$. $r_\text{beam}$ is the laser beam radius at the sample position. 
%\bibliography{RydbergbibJH}

\end{document}